\documentclass[final,3p,times,twocolumn]{elsarticle}

\usepackage{amssymb}
 \usepackage{amsthm}
 \usepackage{amsmath}
 \usepackage{bm}
 \usepackage{adjustbox}
 \usepackage{textgreek}
 \usepackage{soul}
\setstcolor{red}
 \usepackage[version=4]{mhchem}
\usepackage{threeparttable}
\usepackage{colortbl}

\usepackage[intoc]{nomencl}
\usepackage{makeidx}
 \makenomenclature
\usepackage{multicol} 
\usepackage{framed} 
\renewcommand*\nompreamble{\begin{multicols}{2}}
\renewcommand*\nompostamble{\end{multicols}}
\renewcommand{\nomgroup}[1]{%
\ifthenelse{\equal{#1}{G}}{\vspace{2mm}\item[\em{Greek Symbols}]}{
\ifthenelse{\equal{#1}{V}}{\vspace{2mm}\item[\em{Vectors}]}{
\ifthenelse{\equal{#1}{S}}{\vspace{2mm}\item[\em{Subscripts}]}{
\ifthenelse{\equal{#1}{P}}{\vspace{2mm}\item[\em{Superscripts}]}{
\item[\em{Symbols}]}}}}
 }
\usepackage{lineno} 
\usepackage{graphicx}
\usepackage{subfigure}
\usepackage{tabularx}
\usepackage{supertabular,booktabs}

\usepackage[colorlinks=true, plainpages = false, citecolor = blue, 
urlcolor = blue, filecolor = blue, linkcolor=black]{hyperref}
\usepackage{ifthen}
\usepackage[usenames,dvipsnames]{xcolor}
\newcommand{\AM}[1]{{\color{black} #1}}
\usepackage{cprotect}

 \DeclareFontFamily{U}{euc}{}
 \DeclareFontShape{U}{euc}{m}{n}{<-6>eurm5<6-8>eurm7<8->eurm10}{}%
 \DeclareSymbolFont{AMSc}{U}{euc}{m}{n} 
 \DeclareMathSymbol{\umu}{\mathord}{AMSc}{"16}

\journal{International Journal of Heat and Mass Transfer}
\definecolor{lightgray}{gray}{0.95}

\begin{document}


\begin{frontmatter}



\title{\flushleft Experimental measurements of the  permeability of fibrous carbon at high-temperature
}


\author[first]{Francesco Panerai}

\author[fourth]{Jason D. White}

\author[third]{Thomas J. Cochell}
 
\author[first]{Olivia M. Schroeder}

\author[fifth]{Nagi N. Mansour}

\author[sixth]{Michael J. Wright}
 
\author[first]{Alexandre Martin\corref{cor1}}
 \ead{Alexandre.Martin@uky.edu}

\address[first]{Department of Mechanical Engineering, 
University of Kentucky, Lexington, KY 40506, USA}

\address[fourth]{Advanced Technology \& Systems Division, 
SRI International, Menlo Park, CA 94025, USA}
\address[third]{Department of Chemical and Materials Engineering, 
University of Kentucky, Lexington, KY 40506, USA}

\address[fifth]{NASA Advanced Supercomputing Division, 
NASA Ames Research Center, Moffett Field, CA, 94035, USA}

\address[sixth]{Entry Systems and Technology Division, 
NASA Ames Research Center, Moffett Field, CA, 94035, USA}

\cortext[cor1]{Corresponding author.}

\begin{abstract}
A series of experiments was performed to obtain permeability data on
FiberForm\textsuperscript{\textregistered}, a commercial carbon preform used for manufacturing thermal protection systems.  A porous sample was placed in a quartz flow-tube 
heated by an isothermal furnace. 
The setup was instrumented to measure mass flow through and pressure drop across the sample.
The intrinsic permeability and the Klinkenberg correction, which accounts for rarefied effects, were computed from the experimental data.
The role of the gas temperature and pressure on the effective permeability is shown, and it is demonstrated that with proper data reduction, the intrinsic permeability is strictly a function of the micro-structure of the material.
A  function for the effective permeability of  FiberForm, dependent on temperature, pressure, pore geometry, and type of gas is proposed.
The intrinsic permeability was evaluated at $K_0 = 5.57 \times 10^{-11}$ m$^2$, with a Klinkenberg parameter of $8c/d_p = 2.51 \times 10^5$ m$^{-1}$ and a reference porosity of $\phi^\dagger = 0.87$.
 \end{abstract}

\end{frontmatter}

\nomenclature{$u$}{gas velocity [m/s]}
\nomenclature{$A$}{area of the flow-tube [m$^2$]}
\nomenclature{$D$}{diameter of the flow-tube and sample [m]}
\nomenclature{$L$}{length of the sample [K]}
\nomenclature{$m$}{mass of the sample [K]}
\nomenclature{$T$}{temperature [K]}
\nomenclature{$P$}{pressure [Pa]}
\nomenclature{$\Delta P$}{pressure difference across sample [Pa]}
\nomenclature{$\dot{m}$}{mass flow rate [kg/s]}
\nomenclature{$x$}{spatial coordinates [m]}
\nomenclature{$F$}{resistive force [N]}
\nomenclature{$\delta f$}{frequency resolution [s$^{-1}$]}
\nomenclature{$t$}{time [s]}
\nomenclature{$\cal R$}{universal gas constant [J/(K$\cdot$mol)]}
\nomenclature{$M$}{molar mass [kg/mol]}
\nomenclature{$d_p$}{characteristic pore diameter [m]}
\nomenclature{$d_f$}{average fiber diameter [m]}
\nomenclature{$c$}{proportionality constant }
\nomenclature[G]{$\lambda$}{mean free path [m]}
\nomenclature[G]{$\phi$}{porosity [m$^3$/m$^3$]}
\nomenclature{$b$}{permeability slip parameter [Pa]}
\nomenclature{$K_\text{eff}$}{effective permeability [m$^2$]}
\nomenclature{$K_0$}{intrinsic permeability [m$^2$]}
\nomenclature[G]{$\mu$}{viscosity [kg/(m$\cdot$s)]}
\nomenclature[G]{$\rho$}{density [kg/m$^3$]}
\nomenclature[S]{s}{surface}
\nomenclature[S]{f}{furnace}
\nomenclature[S]{1}{port P1}
\nomenclature[S]{2}{port P2}
\nomenclature[S]{avg}{average across sample}
\nomenclature[P]{*}{values scaled at $T=298$ K}
\nomenclature[P]{$\dagger$}{values scaled at $\phi=0.87$}

\begin{table*}[ht!]
\begin{framed}
\printnomenclature
 \end{framed}
\end{table*}


\section{Introduction}
\label{sec:Introduction}

The entry process into a planetary atmosphere requires spacecraft to be equipped with a thermal protection system (TPS). The TPS protects the spacecraft from the high enthalpy and thermochemical conditions of entry, during which the hypersonic flow surrounding the vehicle generates strong aerothermal heating. 
An ablator is usually used as a TPS material for the harshest entry conditions due to the
 chemical and physical phenomena that take place when high heat fluxes are experienced. The ablator materials significantly reduce the heat to the inner parts of the vehicle, protecting the payload~\cite{Bowman1971aa}.  
In recent years the focus 
has veered toward a new class of low-density carbon/resin ablators, the most successful of which is NASA's own phenolic-impregnated carbon ablator (PICA)~\cite{hui}, used in Earth return and Mars exploration missions \cite{Kontinos:2010aa,MSL}.
\begin{figure}[htpb]
\centering
   \includegraphics[width=\columnwidth]{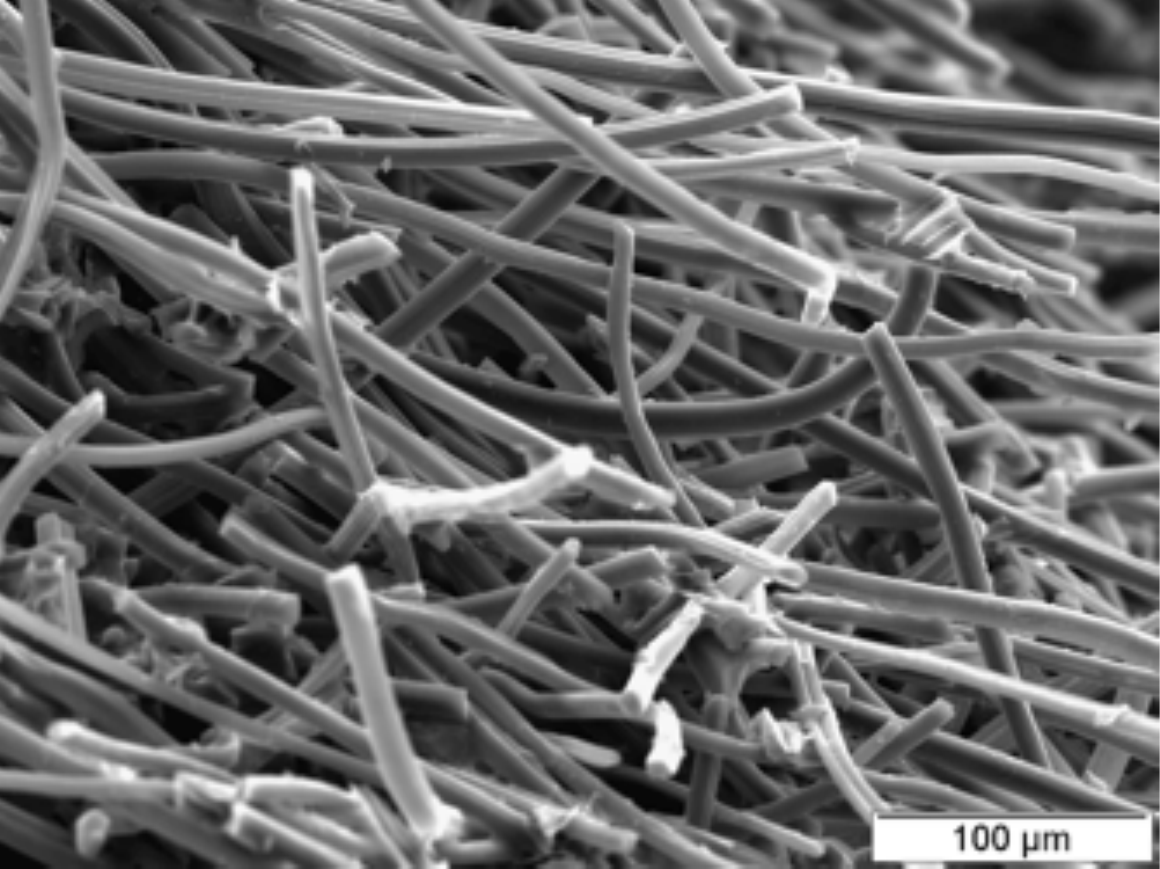}
\caption{Scanning electron micrograph of FiberForm.}
\label{fig1}
\end{figure}

PICA uses FiberForm\textsuperscript{\textregistered} (Fiber Materials, Inc.)~\cite{FiberForm}, a rigid carbon fiber composite, as a substrate. As shown by the scanning electron micrograph in Fig.~\ref{fig1}, its micro-structure is characterized by thin carbon fibers ($\approx$ 10 \textmu m in diameter) and pores of $\approx$ 50 \textmu m in diameter \cite{lachaudJSR}. The pores occupy nearly a 90\% fraction of the volume of the material, providing it with excellent insulation properties.  

Because of their high porosity, gases can easily flow within the ablative materials. For example, pyrolysis gases produced by decomposition of the phenolic resin travels through the charred structure -- potentially reacting with the fibers -- before exiting the material.
Likewise, reactants from the boundary layer can enter the material microstructure and flow within the pores. This gas transport has a significant effect on the overall material response~\cite{jtht2013_weng,ijhmt013_weng,martinJTHT_pyrogas3}. 

The flow behavior through a porous structure is characterized by the permeability, as it dominates the momentum transport within the medium. Permeability is therefore a key material property when modeling porous media flow.

When the mean free path $\lambda$ of the gas molecules approaches the dimensions of the material pores, the gas flow within the material is considered transitional between the continuum and Knudsen regimes. In this regime, slip effects become important.

A method for measuring the permeability of porous refractory insulators was proposed by Marschall and Milos~\cite{marschall2,marschall_cox} and applied to various materials, such as silica-based tiles, PICA (in virgin and charred from), ceramics, and to a lesser extent, FiberForm. Data on FiberForm in Ref.~\cite{marschall2} were obtained up to 300 K on an older, less dense version of the material with large observed variabilities in the samples.

The experiments documented in the current work provide an updated set of FiberForm permeability values at
temperatures ranging from 298 to 1500 K, in inert atmosphere. The data generated also 
constitute a  
baseline for the numerical rebuilding of effective reactivity data from experiments on the high-temperature decomposition of FiberForm \cite{jtht2013_oxidation}.


\section{Experiment}
\label{sec:Experiment}
\begin{figure*}[ht]
\centering
  \includegraphics[width=\textwidth]{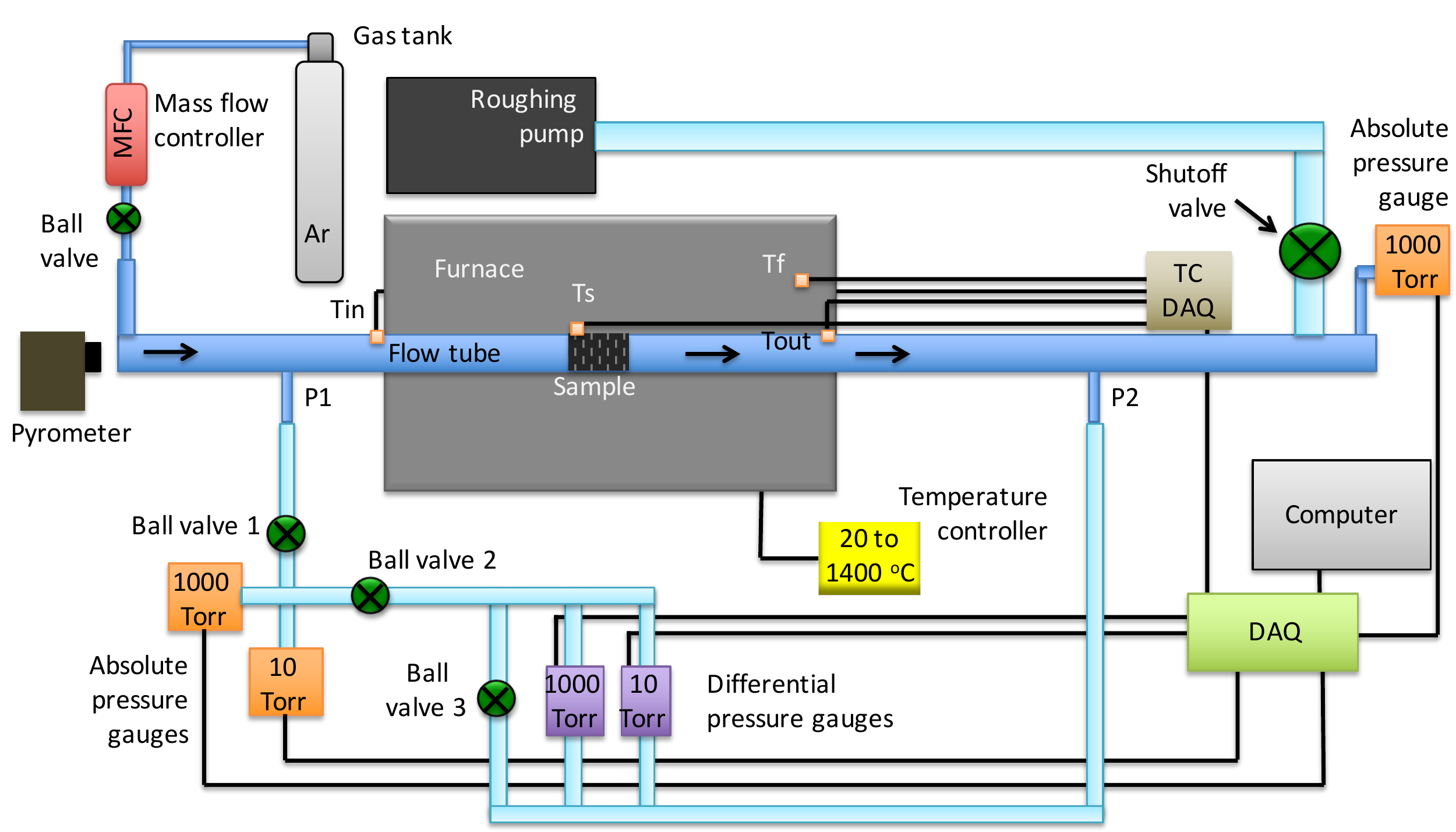}
\caption{Schematic of flow-tube setup at SRI International.}
\label{fig2}
\end{figure*}

A high-temperature flow-tube setup (Fig.~\ref{fig2})  was assembled to perform gas/material interaction experiments on porous samples. 
The system consisted of a 129.5 cm long, 22 mm inner diameter quartz tube positioned inside of an open-ended furnace providing temperatures up to 1675~K by means of a radiative ceramic element. The cylindrical plug samples (FiberForm) were inserted in the tube by interference fitting, and positioned in  the center of the furnace, using a plastic dowel rod.
As discussed in the related 
literature \cite{marschall2,marschall_cox}, the axial geometry of the porous material plays a major role in the permeability.

\AM{Because of its manufacturing method, FiberForm is an orthotropic material. More specifically, it is transverse isotropic, since most of the fibers are oriented  within  $\pm$15$^{\circ}$ of the compression plane. The direction perpendicular to this plane is defined as ``Through-Thickness" (TT) and that parallel as ``In-Plane" (IP).}
The bulk of the experiments described here were performed on samples machined with a TT orientation, in which the carbon fibers are preferentially aligned perpendicular to the gas flow direction.
One experiment was also performed with a sample oriented in the IP direction.
A dedicated mass flow controller (Aalborg Model: UFC 8160) calibrated to nitrogen trifluoride was controlled by a Tylan RO-28 Readout/Control Box to feed the argon gas at fixed flow rates ranging between 10 and 100 sccm. The system was evacuated by means of an Alcatel R301B Roots pump using Fomblin\textsuperscript{\textregistered} oil and backed by an Alcatel BF ADP 81 dry pump. The pumping manifold was outfitted with a copper mesh to collect particulates that might be emitted during the experiment. The outlet of the flow-tube was connected to the vacuum system through a manual bellows-angle valve, fully opened during the experiments.

The main tube was equipped with both an upstream (P1) and downstream (P2) port from the furnace, which were connected to a manifold of calibrated differential pressure transducers measuring the pressure loss  ($P_1-P_2$) across the sample. A separate set of pressure gauges were also used to monitor absolute pressure conditions.

 A valving manifold was used to control gas flows and normalize pressure in the system when starting experiment operations. One set of these valves was used as a by-pass to prevent the formation of strong pressure gradients across the sample during evacuation or venting operations that could potentially move the sample from the desired initial position. Thermocouple (TC) sensors were installed at different strategic positions along the tube as depicted in Fig.~\ref{fig2}. Two Type-K thermocouples were used to measure the temperature $T_1$ and $T_2$ 
 at the pressure ports P1 and P2, respectively, and two other Type-K TCs monitored the temperature $T_\text{in}$ and $T_\text{out}$ at the inlet and outlet of the furnace. A Type-S and Type-K TC were used to measure the temperature of the sample $T_s$ and the temperature of the furnace $T_f$, respectively, at the position of the carbon plug. 
\AM{A two-color pyrometer (Mikron M90-R2) pointed at the upstream surface of the sample was also used as a redundant temperature measurement for the tests at temperature above 1200 K. All of these temperature measurements agreed within 10 K.}

Pressure and temperature measurements were acquired at 4 s intervals by a dedicated acquisition card (NI Model USB-6210). A customized LabView interface recorded each parameter directly to a computer.



The protocol for the experiments followed that of Marschall and Milos~\cite{marschall2}.
First, length and mass of the FiberForm samples were measured, and sample densities were calculated. The samples were then inserted into the Quartz tube, the system was evacuated to a base pressure $P < 13.33$ Pa, and the background temperature was stabilized to a steady target while supplying a constant 10 sccm Ar flow to the system. 


The permeability of virgin char in an \ce{Ar} environment was measured as follows. 
Absolute pressure, differential pressure, and temperature data were collected at a frequency resolution of $\delta f=0.25$~Hz until a steady state was reached. 
Gas flow was stopped following the measurement.

The furnace was then cooled, and the FiberForm samples were removed by pushing from the backside of the plug toward the pyrometer. Post-testing mass and length measurements were performed on the samples. Negligible mass loss and length changes were measured compared with pre-testing measurements.

\section{Permeability in the slip regime}

The permeability of FiberForm was determined by measuring the pressure gradient $\Delta P$ across the sample for a given combination of temperature $T$, mass flow rate $\dot{m}$, and gas mixture including viscosity $\mu$ and molar mass $M$. 
Klinkenberg derived an expression that accounts for non-continuum effects in porous media~\cite{klinkenberg}. This equation describes the permeability as a function of the Knudsen number $\text{Kn} = \lambda/d_p$. Here, $\lambda$ is the mean free path of gas molecules and $d_p$ is the mean pore diameter of the material, assumed to be the characteristic length of the porous medium. This Klinkenberg expression takes the form of an effective permeability
\begin{equation}
K_{\text{eff}} = K_0\left(1+ 8c  \frac{\lambda}{d_p}\right)
\label{eq1}
\end{equation}
where $K_0$ is the value of the permeability in the limit of continuum flow regime,  
and $c$ is a proportional constant. Accounting for pressure and temperature, the mean free path can be expressed as: 
\begin{equation}
\lambda=  \frac{\mu(T)}{P}  \sqrt{\frac{\pi}{2}\frac{ {\cal R} T}{M}}
\label{mean_free_path}
\end{equation}
To simplify the notation, it is convenient to define parameter $b(T)$ according to
\begin{equation}
b(T) = \frac{8c}{d_p} \mu(T)  \sqrt{\frac{\pi}{2}\frac{ {\cal R} T}{M}}
\label{b}
\end{equation}
thus obtaining the Klinkenberg expression for the effective permeability:
\begin{equation}
K_\text{eff} = K_0\left( 1+ \frac{b}{P}\right)
\label{kpb}
\end{equation}
%
Because the flow-tube configuration produces a well-defined and well-characterized flow, it is possible to analyze the flow field using an analytical approach and extract the permeability parameters from the experimental results.
By combining Eq.~\ref{kpb} with the conservation of mass, ideal gas law, the geometry of the sample, and Darcy's Law, the following relationship can be derived for one dimensional, laminar and isothermal flows: 

%
%
\begin{equation}
 \dot{m} = - \rho u  A =  - \left[\frac{PM}{{\cal R}T} \right]      \left[\frac{  \pi D^2}{4}\right]  \left[ \frac{K_\text{eff}}{\mu} \frac{dP}{dx}\right]   
 \label{eq2}
\end{equation}
\begin{equation}
\frac{4 \mu \dot{m} {\cal R} T}{ \pi D^2 M} \int_{0}^{L}dx = - K_0 \int_{P_1}^{P_2}(P + b) dP
\label{eq3}
\end{equation}
\begin{equation}
F = \frac{4 \mu  \dot{m} {\cal R} TL}{ \pi D^2 M \Delta P} = K_0(P_\text{avg} + b)
\label{eq4}
\end{equation}

In these equations, $L$ and $D$ are length and diameter of the sample, and $P_{avg}=0.5(P_1+P_2)$ is the average pressure in the sample. As for $F$, it is a force that results from the  material permeability, and is only dependent on known constants or measured experimental values.

For the conditions of the current experiments, the argon flow within the porous medium is in the rarefied regime. For temperatures between 1000 and 2000 K, and pressures between 1 and 10 kPa, the Knudsen number remains between 0.06 and 2. This wide variation supports the use of the Klinkenberg equation to correct the permeability. 

\section{High-temperature permeability measurements \label{sec:section4}}

Since all  quantities in the left-hand side of Eq.~\ref{eq4} are known or measured,  $K_0$ and $b$ can be obtained by a linear least-squares fit of $F=F(P_{avg})$. The slope provides the value for  $K_0$, and  depends on the material micro-structure only. The abscissa at zero ordinate divided by $K_0$ provides $b$ and depends on the flow temperature, the type of gas, as well as on the material micro-structure through the average pore diameter.

\begin{figure}[htpb]
\centering
   \includegraphics[width=\columnwidth]{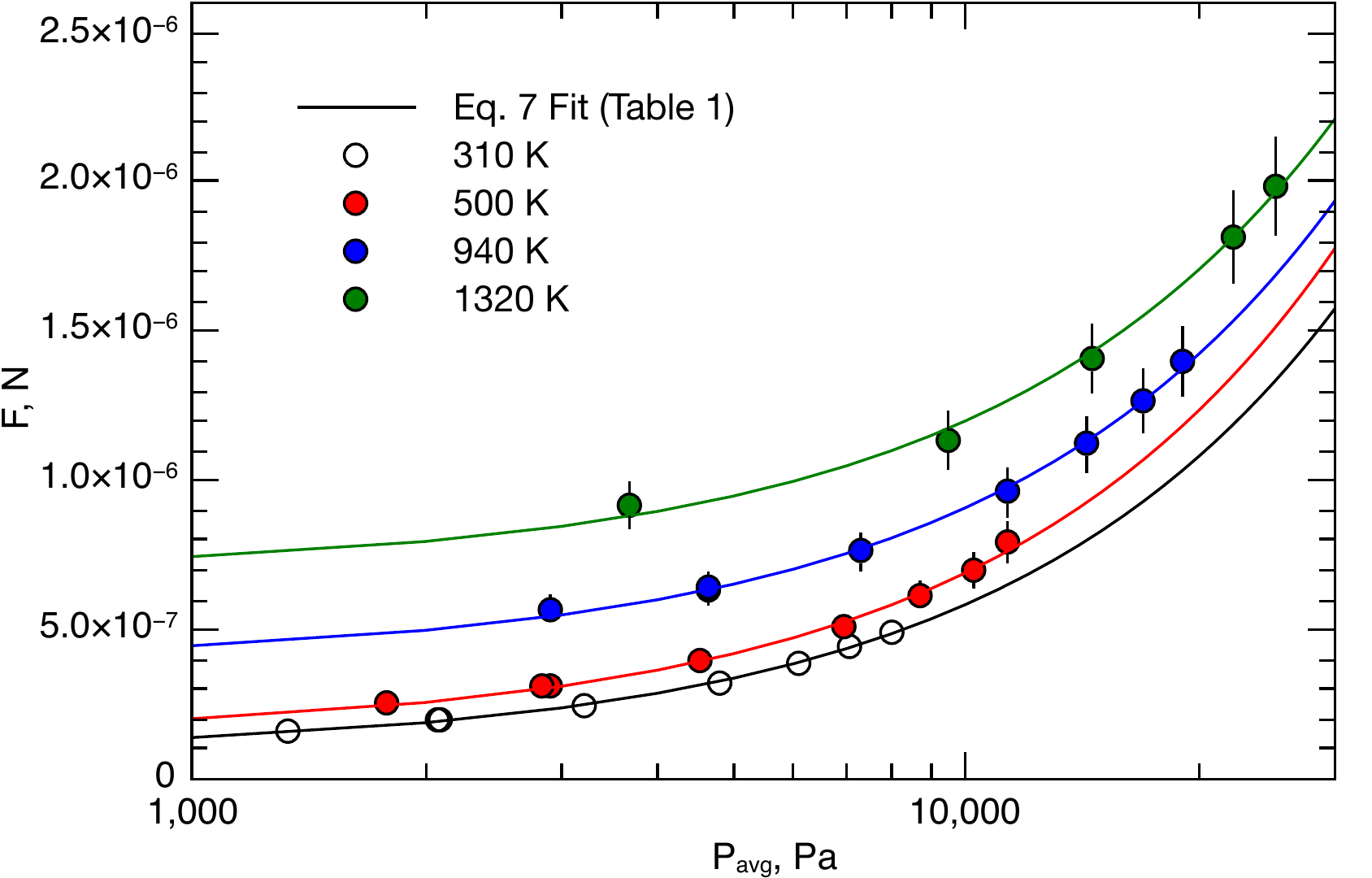}
\cprotect\caption{Measured permeability values in \ce{Ar} flow for sample \verb+TT07+, as a function of temperatures.}
\label{graph1}
\end{figure}

An example of permeability measurements in \ce{Ar} flow, at temperatures from 310 to 1320 K
, is shown in Fig.~\ref{graph1} for sample \verb+TT07+. 
$F$ is a linear function of $P$ with constant slope. To build a representative database for the 
permeability of FiberForm, multiple samples were tested at various conditions. The results from these tests are presented in Table~\ref{free3}. The full set of measured data is provided in Table~\ref{exp_data} of the supplementary material.

\begin{center}
\begin{threeparttable}[htpb]
\caption{Temperature-dependent permeability data}\label{free3}
\begin{small}
\begin{tabular}{l crcr  }
\toprule
Sample 	&	$\rho$, kg/m$^3$	 &	$T$, K  &	$K_0$, m$^2$	&	$b$, Pa	\\
\midrule
\verb+TT01+ 	 &	192  &	298	&	$4.97\times 10^{-11}$	&	1620	\\
							&&	723	&	$5.50\times 10^{-11}$	&	3850	\\
							&&	1123	&	$5.34\times 10^{-11}$	&	8360	\\
							&&	1503	&	$5.91\times 10^{-11}$	&	12600	\\ 
							
\rowcolor{lightgray}
\verb+TT02+ & 	187 	&	297	&	$6.07\times 10^{-11}$	&	1450	\\
\rowcolor{lightgray}
							&&	933	&	$6.16\times 10^{-11}$	& 	5990	\\
\verb+TT03+ &	182	&	361	&	$6.24\times 10^{-11}$	&	1980	\\
							&&	1121	&	$5.99\times 10^{-11}$	&	8530	\\
\rowcolor{lightgray} 
\verb+TT04+ &	181	&	297	&	$5.46\times 10^{-11}$	&	1670	\\
\rowcolor{lightgray}
							&&	723	&	$5.38\times 10^{-11}$	&	4660	\\
\verb+TT05+ &	189	&	391	&	$5.41\times 10^{-11}$	&	2310	\\
							&&	823	&	$5.14\times 10^{-11}$	&	5580	\\
\rowcolor{lightgray} 
\verb+TT06+ &	178	&	297	&	$6.05\times 10^{-11}$	&	1590	\\
\rowcolor{lightgray}				
							&&	523	&	$6.09\times 10^{-11}$	&	2760	\\

\verb+TT07+ &	186	&	310	&	$4.95\times 10^{-11}$	&	1810	\\
							&&	503	&	$5.43\times 10^{-11}$	&	2800	\\
							&&	940	&	$5.14\times 10^{-11}$	&	7730	\\
							&&	1320&	$5.06\times 10^{-11}$	&	13700	\\

\rowcolor{lightgray} 
\verb+TT08+ &	177	&	298	&	$6.07\times 10^{-11}$	&	2070	\\
\rowcolor{lightgray}
							&&	523	&	$6.72\times 10^{-11}$	&	2850	\\
\rowcolor{lightgray}
							&&	731	&	$5.71\times 10^{-11}$	&	6790	\\
\rowcolor{lightgray}
							&&	935	&	$5.75\times 10^{-11}$	&	8900	\\
\rowcolor{lightgray}
							&&	1130	&	$5.96\times 10^{-11}$	&	10515	\\
\rowcolor{lightgray}
							&&	1321	&	$5.94\times 10^{-11}$	&	13530	\\
\rowcolor{lightgray}
							&&	1507	&	$5.99\times 10^{-11}$	&	17559	\\

\verb+TT09+ &	181	&	297	&	$5.46\times 10^{-11}$	&	3070	\\
							&&	1421	&	$6.59\times 10^{-11}$	&	10100\\
\rowcolor{lightgray}
\verb+IP01+\tnote{1}   &     186    &    298     &      $1.12\times 10^{-10}$	&      1408\\
\rowcolor{lightgray}
							&&   723   &     $1.13\times 10^{-10}$         &	4348\\
\rowcolor{lightgray}
							&&   1123 &     $1.15\times 10^{-10}$	         &     8176\\
\rowcolor{lightgray}			
							&&   1503 &     $1.24\times 10^{-10}$	 &    12810\\
\bottomrule
\end{tabular}
\end{small}
\footnotesize 
\begin{tablenotes}
         \item[1] \scriptsize In-Plane orientation: all other samples are Through-Thickness
\end{tablenotes}
\end{threeparttable}
\end{center}

\begin{figure}[htpb]
\centering
   \includegraphics[width=\columnwidth]{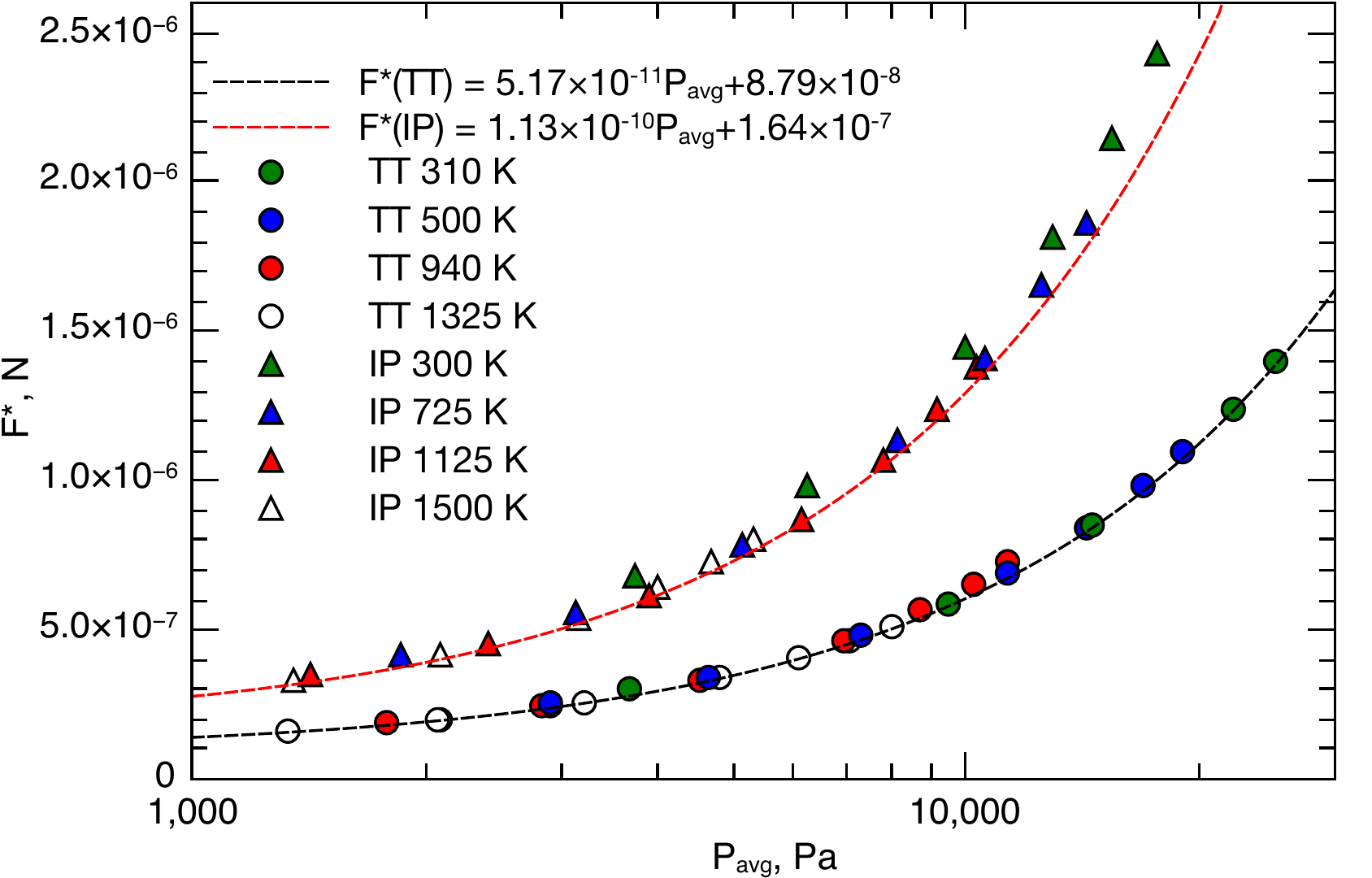}
\cprotect\caption{Normalized permeability data for
sample \verb+TT07+ (Through-Thickness) and 
sample \verb+IP01+ (In-Plane) highlighting the transverse orthotropic properties of FiberForm. Fitting curves are displayed in black and red for Through-Thickness and In-Plane orientation, respectively.}
\label{graph2}
\end{figure}

The dependency of parameter $b$ on the temperature and the type of gas can be removed by normalizing the data to a standard condition for the gas, here chosen to be 298 K. Replacing temperature-dependent terms with an asterisk, Eq.~\ref{eq4} becomes $F^* = K_0 (P_{avg}+b^*)$, where $b^*$ is calculated by normalizing Eq.~\ref{b} as
\begin{equation}
\frac{b}{b^*}= \frac{\mu}{\mu^*}  \sqrt{\frac{\pi}{2} \frac{{\cal R}T}{M}}\sqrt{\frac{2}{\pi}\frac{M^*}{{\cal R}T^*}}
\label{bobs}
\end{equation}
Here, $\mu^*$ is the viscosity for \ce{Ar} at reference temperature $T^*$ = 298 K. The molar mass ratio $M^*/M$ is equal to 1, since only \ce{Ar} is used. Using the process, the curves of Fig.~\ref{graph1} collapse onto a single curve shown with circles in Fig.~\ref{graph2}.

\AM{New curve fits can be calculated for the scaled data, and the  $8c/d_p$  term in Eq.~\ref{eq1} can be calculated for each sample. These values of  $8c/d_p$  are listed in Table~\ref{results2}.} 

Fig.~\ref{graph2} also illustrates the strong transverse isotropic properties of FiberForm with a comparison to the IP configuration. This geometry allows the gas to flow more easily along the axial direction of the planar alignment of the fibers, and results into a higher permeability than in the TT direction. Both curves have different slopes, which results into distinctive intrinsic permeability values $K_0$. The numerical values, presented in Table~\ref{results2}, also show that the rarefied term, $8c/d_p$ does not vary with direction, as expected, since it is only a function of the average pore size.

Table~\ref{results2} shows a non-negligible scatter in the measured density of the samples due to the method of fabrication of FiberForm that generates inhomogeneities, as can be seen in Fig.~\ref{fig1}.
To further normalize the samples, $F^*$ becomes $F^\dagger$ using a factor ${\phi^\dagger}/{\phi}$ that accounts for the deviation of the density of each sample from the nominal density.

\begin{figure}[htpb]
\centering
   \includegraphics[width=\columnwidth]{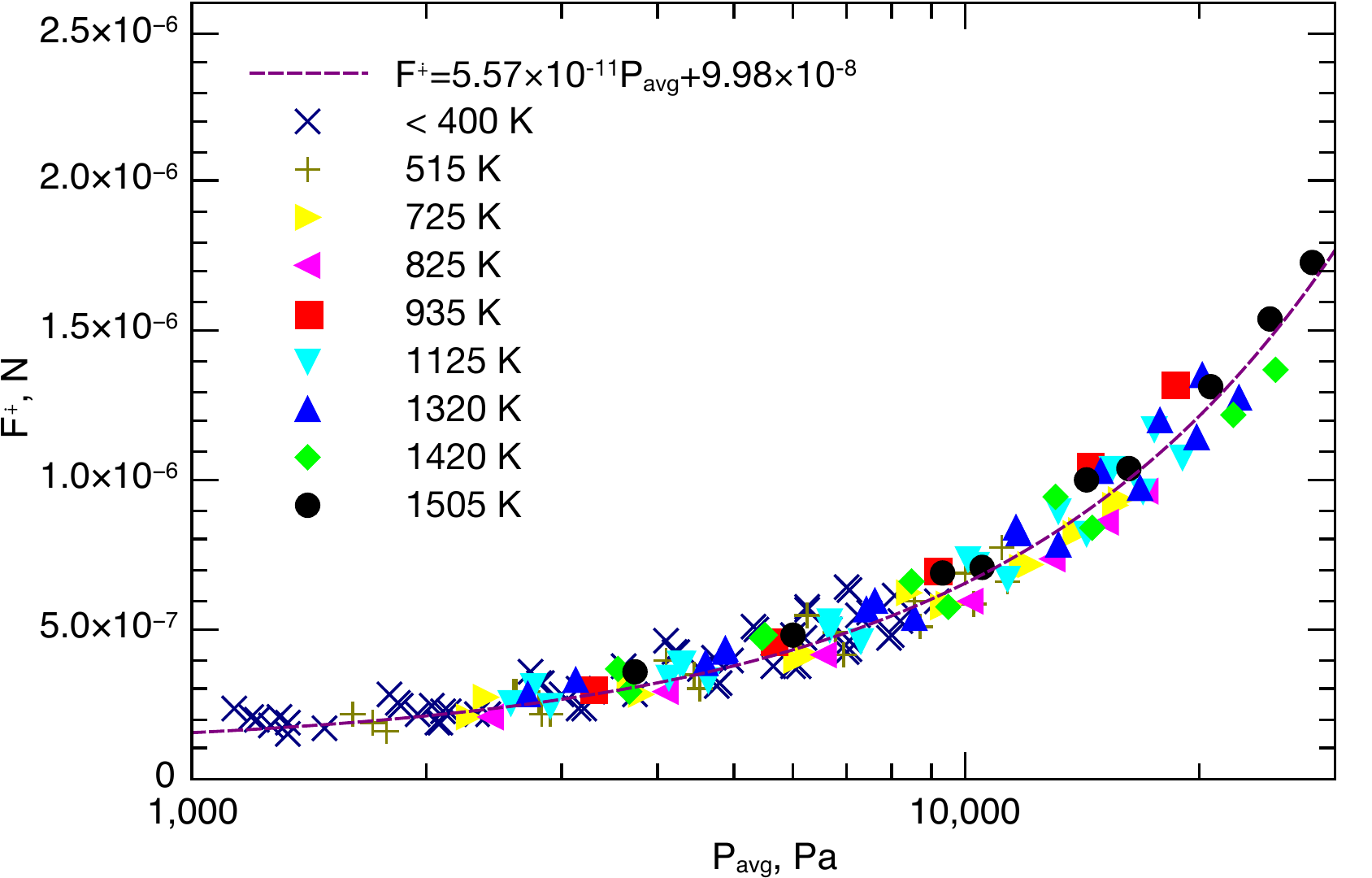}
\caption{Summary of FiberForm permeability measurements.}
\label{fig4}
\end{figure}

In this factor, the porosity $\phi$ is calculated according to:
\begin{equation}
\phi = 1 - \frac{\rho}{\rho_C}
\end{equation}
where $\rho_C$ is the density of the fibers. A value of $\rho_C =1400$ kg/m$^3$ is calculated by using the average density of the samples $\rho^\dagger = 183.6$ kg/m\textsuperscript{3} and the average reported open porosity of FiberForm is $\phi^\dagger = 0.869$~\cite{FiberForm}.
The new normalized results using  $F^\dagger$ are plotted in Fig.~\ref{fig4}.

Despite this density normalization, some scatter remains, 
likely due to the non-uniformity of the specific micro-structure of each sample, or the Forchheimer effects, which account for inertial deviations to Darcy's law at high velocities~\cite{martinJTHT2009-1,zeng}. 
As was also the case for other classes of porous ablators \cite{marschall2}, the scatter is most pronounced at high mass flow rates, reinforcing the Forchheimer effect hypothesis. Nevertheless, a parametric curve can be fitted to the data, and a single expression for the permeability, based on the normalization and Eqs.~\ref{eq1} and \ref{mean_free_path}, can be obtained:

\begin{equation}
K_{\text{eff}} = K_0\left(1+    \frac{8c}{d_p}  \frac{\mu(T)}{P}  \sqrt{\frac{\pi}{2}\frac{ {\cal R} T}{M}} \right) \frac{\phi}{\phi^\dagger}
\label{eqfubal}
\end{equation}
where $K_0 = 5.57~\times~10^{-11}$ m$^2$,  $8c/d_p = 2.51 \times 10^5$ m$^{-1}$ and $\phi^\dagger = 0.87$. Equation~\ref{eqfubal} can therefore be directly used in Material Response codes when modeling FiberForm, and is valid at any temperature and pressure, in the range covered in the experiment. 

The proportionality constant $c$ can be evaluated using the mean pore diameter for the FiberForm fibrous structure obtained from the porosity. Eichhorn \cite{Eichhorn2005aa} gives the following expression for the average In-Plane pore diameter of a fibrous 2D material:
\begin{equation}
 d_{p,\text{IP}} = \frac{\sqrt{\pi}}{2} \left( \frac{\pi}{2 \ln(1/\phi)} - 1 \right) d_f 
 \label{IPpore_diam}
\end{equation}
Using an average fiber diameter $d_f$ of 11 $\umu$m, a value of $d_{p,\text{IP}}\approx96$ $\umu$m is estimated.
From Ref.~\cite{sampson2008contribution}, In-plane pore diameter $d_{p,\text{IP}}$ and pore height $\bar{h}$, for cylindrical fibers, can be related using the following expression:
\begin{equation}
 \frac{\bar{h}}{d_{p,\text{IP}}}=\frac{\sqrt{\phi}}{2}
 \label{pore_height}
\end{equation}
from which $\bar{h}\approx$ 45 $\umu$m.
Therefore, with $d_p=\bar{h}$, a value of $c=1.6$ is obtained.

\begin{figure}[htpb]
\centering
   \includegraphics[width=\columnwidth]{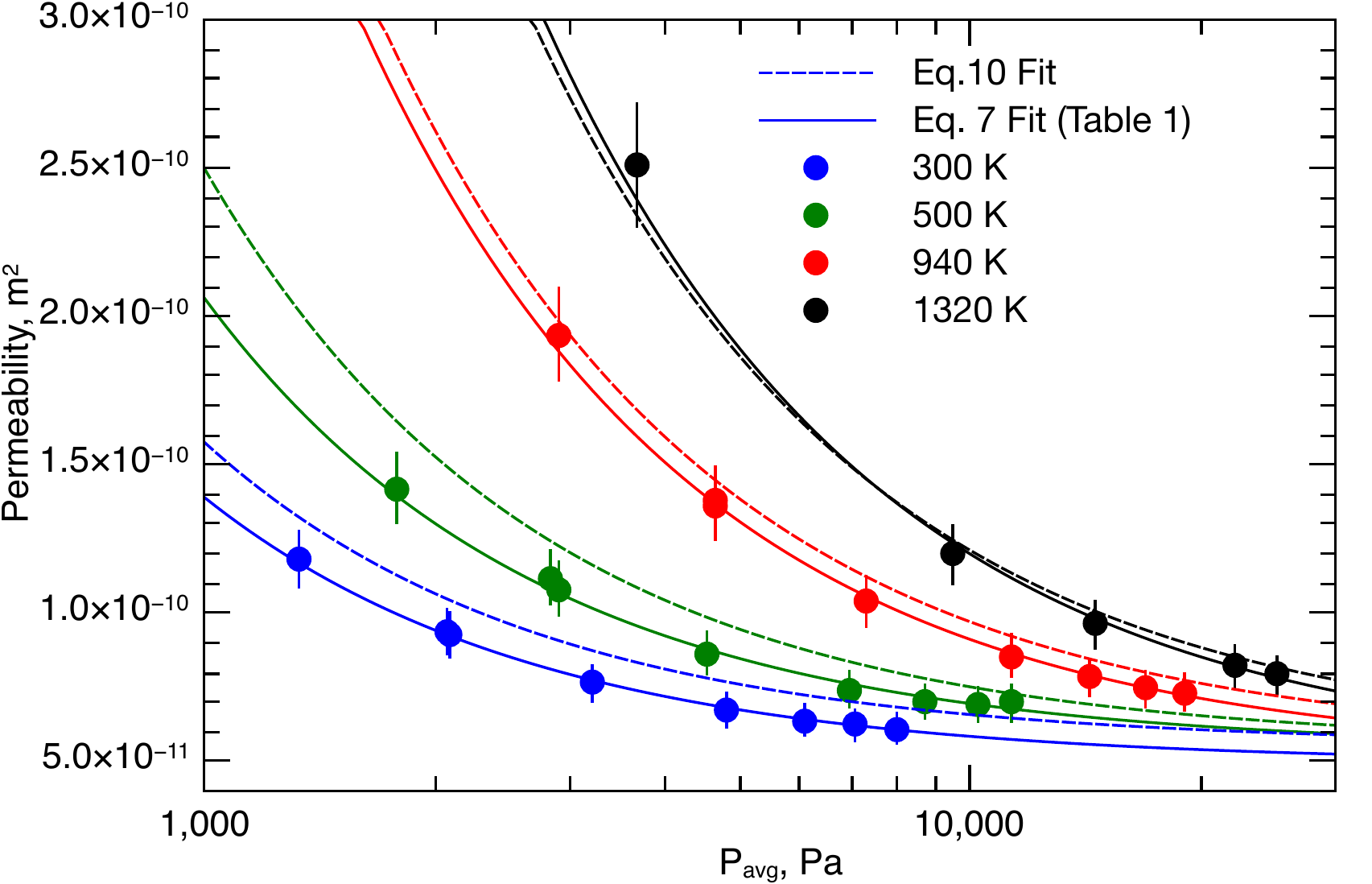}
\caption{Comparison  permeability values obtained using Eq.~\ref{eqfubal},  Eq.~\ref{kpb}, and the experimental data.}
\label{perm}
\end{figure}

Figure~\ref{perm} compares the experimental results of the permeability with the values obtained using Eq.~\ref{eqfubal} for sample \verb+TT07+. While good agreement is shown, there is no perfect match, since Eq.~\ref{eqfubal} uses all the dispersed experimental data to generate the curve fit.
The permeability values obtained using Eq.~\ref{kpb}, with the parameters listed in Table~\ref{free3}, 
are also plotted in the same figure. The fit based on Eq.~\ref{kpb} is  closer to the experimental values since it was generated using these.
Outside of the pressure range where measurements were collected, the fit is less accurate but still within acceptable errors, and the use of Eq.~\ref{eqfubal} is recommended for modeling the permeability in numerical simulations.

\definecolor{lightgray}{gray}{0.95}
\begin{center}
\begin{threeparttable}[htpb]
\caption{Normalized permeability data}\label{results2}
\begin{small}
\begin{tabular}{l ccc  }
\toprule
Sample 	&	$\rho$, kg/m$^3$	 &	$K_0$, m$^2$	&	$8c/d_p$, m$^{-1}$	\\
\midrule
\verb+TT01+ &	192 	&	$5.43\times 10^{-11}$	&	$2.12\times 10^{5}$	\\
\rowcolor{lightgray}
\verb+TT02+ & 	187 	&	$6.18\times 10^{-11}$	&	$1.96\times 10^{5}$	\\
\verb+TT03+ &	182	&	$5.96\times 10^{-11}$	&	$2.48\times 10^{5}$	\\
\rowcolor{lightgray} 
\verb+TT04+ &	181	&	$5.31\times 10^{-11}$	&	$2.49\times 10^{5}$	\\
\verb+TT05+ &	189	&	$5.08 \times 10^{-11}$	&	$2.66\times 10^{5}$	\\
\rowcolor{lightgray} 
\verb+TT06+ &	178	&	$5.98\times 10^{-11}$	&	$2.22\times 10^{5}$	\\
\verb+TT07+ &	186	&	$5.17\times 10^{-11}$	&	$2.37\times 10^{5}$	\\
\rowcolor{lightgray} 
\verb+TT08+ &	177	&	$5.60\times 10^{-11}$	&	$2.84\times 10^{5}$	\\
\verb+TT09+ &	181	&	$6.26\times 10^{-11}$	&	$2.98\times 10^{5}$	\\
\rowcolor{lightgray}
\verb+IP01+\tnote{1}  &     186    &    $1.13\times 10^{-10}$	&      $2.05\times 10^{5}$\\
\bottomrule
\end{tabular}
\end{small}
\footnotesize
\begin{tablenotes}
         \item[1] \scriptsize In-Plane orientation: all other samples are Through-Thickness
\end{tablenotes}
\end{threeparttable}
\end{center}

\section{Error Analysis}

The uncertainty associated with the calculation of $F$ depends on the uncertainty contained in the measured and calculated parameters found in Eq.~\ref{eq4}, which are $x_\text{i} = \{ \mu, \dot{m}, R, T, L, D, M, \Delta P\}$. If the uncertainty given by the variables $\delta x_\text{i}$ is small and there is no covariance between them, the error contained in $F$ can be written as:

\begin{equation}
\frac{\delta F}{F} = \left[\sum_{i=1}^{N} \left(\frac{\partial F}{\partial x_{i}}\right)^2 \delta_{x_{i}}^2 \right]^\frac{1}{2}
\label{eq5}
\end{equation}

It should be noted that, since $\mu$ and $T$ are dependent variables, the null covariance condition for the Taylor series expansion is not strictly satisfied; however, it is estimated that considering $\mu$ and $T$ as independent variables does not significantly affect the values of the calculated uncertainty. 
In order to simplify Eq.~\ref{eq5}, $F$ can be expressed as:

\begin{equation}
F=\prod_{i=1}^{N} x_{i}^{\alpha_{i}}
\label{eq6}
\end{equation}

\begin{equation}
\ln{F} = \sum_{i=1}^{N}\alpha_{i}\ln x_{i}
\label{eq7}
\end{equation}

Thus, the variance of $F$ becomes:

\begin{equation}
\delta_{\ln{F}}^{2} = \sum_{i=1}^{N} \alpha_{i}^{2}\delta_{\ln{x_{i}}}^{2}
\label{eq8}
\end{equation}

If sufficiently small ($\leq$ 20 \%), the standard deviation of the natural logarithm of a random variable is approximately equal to the relative standard error, i.e., $\delta_{\ln{x_{i}}} = {\delta_{x_{i}}}/{x_{i}}$. Therefore, Eq.~\ref{eq5} can be approximated with sufficient accuracy as:

\begin{equation}
\frac{\delta F}{F}=\left[\sum_{i=1}^{N}\alpha_{i}^{2}\left(\frac{\delta_{x_{i}}}{x_{i}}\right)^2\right]^\frac{1}{2}
\label{eq9}
\end{equation}

Expanding the series for all parameters, the expression becomes:
\begin{equation}
\begin{split}
\frac{\delta F}{F}  = \ & \Biggl[\left(\frac{\delta_\mu}{\mu}\right)^2+\left(\frac{\delta_{\dot{m}}}{\dot{m}}\right)^2+\left(\frac{\delta_R}{R}\right)^2+\left(\frac{\delta_T}{T}\right)^2\\
&+\left(\frac{\delta_L}{L}\right)^2+\left(\frac{\delta_D}{D}\right)^2+\left(\frac{\delta_M}{M}\right)^2+\left(\frac{\delta_{\Delta_P}}{\Delta P}\right)^2\Biggr]^\frac{1}{2}
\end{split}
\label{eq10}
\end{equation}

In Eq.~\ref{eq10}, $\delta_R$ and $\delta_M$ are negligible, and $\delta_L$ and $\delta_D$ are very small. Therefore, the main contributions to errors on $F$ are associated with mass flow rate, temperature, and consequently gas viscosity and pressure measurements. A list of the determined uncertainties associated with each input variable is provided in Table 1. The total uncertainty associated with $F$ was calculated using Eq.~\ref{eq10} to be $\pm$ 8.2 $\%$.

One additional source of error in $K_0$ and $b$ is due to the least-square fit of the $F=F(P_{av})$ function. For the experiments presented above, the coefficient of determination was found to be $R^2>0.99$ for each of the acquired measurements. Using the parameters in Table~\ref{uncertainty}, by numerical fits, the uncertainty associated with $K_0$ was found to be $\pm 10\%$ and with $b$ was found to be $\pm 2\%$.

\begin{table}[ht!]
\caption{Determined uncertainty of experimental parameters.}\label{uncertainty}
\begin{center}
\begin{small}
\begin{tabular}{ cc}
\toprule
Observed Variable 	&	Uncertainty ($\%$)  \\
\midrule
$\mu$         	&	$\pm$ 3	\\
$\dot{m}$    	&	$\pm$ 7	\\
$R$     		&	$\pm$ 0.1	\\
$T$   		&	$\pm$ 2	\\
$L$    		&	$\pm$ 1	\\
$D$     		&	$\pm$ 0.5	\\
$M$     		&	$\pm$ 0.1	\\
$\Delta P$		&	$\pm$ 2	\\
$F$			&	$\pm$ 8.2	\\
\bottomrule
\end{tabular}
\end{small}
\end{center}
\end{table}

A final remark on the overall uncertainty in permeability is related to the large-scale variability of FiberForm. As shown in Section \ref{sec:section4}, small material samples show a scatter in density on the order of 10\%. The experiment presented here captures the dependence of local permeability parameters with local density. However, visual inspection of larger samples reveals density variations in the Through-Thickness direction, more specifically compressed and expanded layers, and also wavy areas where the local Through-Thickness vector varies in orientation, perhaps by 10 of 15 degrees. It may be appropriate to consider such large-scale variabilities when modeling a large volume of FiberForm.

\section{Conclusions}
The presented work provides a benchmark dataset of the permeability properties of FiberForm. This approach is fundamental in that it allows for a wide variety of experimental testing on internal oxidation reactions in partially dissociated air flows and the chemical evolution of representative pyrolysis gas mixtures by heterogeneous surface reactions with a hot char. 
Furthermore, a standard practice has been defined for handling porous materials in a flow-tube, which can be applied to oxidation experiments with reactive gas species. In our protocol, the acquisition of permeability data at both room temperature and at oxidation temperature, prior to the start of the reactive gas/material interaction and material recession phase, can be used as a tool for comparison to verify the absence of leakage in the interference fitting of the sample in the tube. 

Finally, by defining temperature- and pressure-independent permeability parameters, the work presented here provides an improvement to the material property databases used in high-fidelity Computational Fluid Dynamics and Material Response codes, which are of the utmost importance as the need for more accurate spacecraft re-entry simulations increases.

\section{Conflict of interest}
None declared.

\section*{Acknowledgments}
Financial support for this work was provided by NASA Award NNX14AI97G. The authors are grateful J. Marschall for initiating this project, as well as to  F. S. Milos and Y.-K. Chen for 
reviewing the manuscript and providing constructive comments.



\bibliography{fiber_short_communication}
\clearpage
\onecolumn
\center
\section*{Supplementary Material}

\begin{small}
\tablefirsthead{
\toprule
Sample 	&	$\rho$, kg/m$^3$	 & L, mm & $T$, K  &	$\dot{m}$, kg/s	&	$\Delta P$, Pa &	$P_\text{avg}$, Pa	\\
\midrule
}
\tablehead{%
\multicolumn{2}{c}%
{{\bfseries  Continued from previous page}} \\
\toprule
Sample 	&	$\rho$, kg/m$^3$	 & L, mm &$T$, K  &	$\dot{m}$, kg/s	&	$\Delta P$, Pa &	$P_\text{avg}$, Pa	\\ 
\midrule}
\tabletail{%
\midrule \multicolumn{6}{r}{{Continued on next page}} \\ }
\tablelasttail{%
\bottomrule}
\topcaption{Experimental Measurements.}\label{exp_data}
\begin{supertabular}{lrrrrrr}
\verb+TT01+ 	 &	192  & 20.1 & 298	& 1.8021$\times 10^{-5}$ & 5731.4 &3159.0\\
									 &  &  &  &  3.5962$\times 10^{-5}$ &8614.3 &4750.2\\
									 &  &  &  &  5.4082$\times 10^{-5}$ &10821.9 &5992.5\\
									 &  &  &  &  7.2113$\times 10^{-5}$ &12692.2 &7028.0\\
									 &  &  &  &  9.0143$\times 10^{-5}$ &14323.4 &7957.5\\
							&&&	723	&	1.8021$\times 10^{-5}$ & 11477.5 & 6219.8\\
									 &  &  &  &  3.6052$\times 10^{-5}$ & 17788.9 & 9512.3\\
									 &  &  &  &  5.4172$\times 10^{-5}$ & 22682.4 & 12093.1\\
									 &  &  &  &  7.2203$\times 10^{-5}$ & 26432.3 & 14102.6\\
									 &  &  &  &  9.0143$\times 10^{-5}$ & 29569.6 & 15806.4\\
							&&&	1123	& 1.8112$\times 10^{-5}$ & 15324.3 &8585.1\\
									 &  &  &  &  3.6142$\times 10^{-5}$ & 24295.5 & 13225.4\\
									 &  &  &  &  5.4263$\times 10^{-5}$ & 31249.4 & 16851.0\\
									 &  &  &  &  7.2293$\times 10^{-5}$ & 37029.6 & 19903.5\\
									 &  &  &  &  9.0143$\times 10^{-5}$ & 41995.9 & 22488.7\\
							&&&	1503	& 1.8103$\times 10^{-5}$ & 18365.1  & 10509.6\\
									 &  &  &  &  3.6133$\times 10^{-5}$ &	29505.3 &	16266.9\\
									 &  &  &  &  5.4163$\times 10^{-5}$	&38264.6	 &20804.4\\
									 &  &  &  &  7.2284$\times 10^{-5}$	&45683.7	&24669.7\\
									 &  &  &  &  9.0134$\times 10^{-5}$	&52071.4	&28019.4	\\ 
							
\rowcolor{lightgray}
\verb+TT02+ & 	187 	& 20.1& 297	&	4.5887$\times10^{-6}$ & 2105.6 & 1214.0\\
\rowcolor{lightgray}
									 &  &  &  &  9.0963$\times10^{-6}$ & 3328.9 & 1867.6\\
\rowcolor{lightgray}
									 &  &  &  &  1.8021$\times10^{-5}$ & 5141.4 & 2840.4\\
\rowcolor{lightgray}
									 &  &  &  &  2.7127$\times10^{-5}$ & 6545.3 & 3602.7\\
\rowcolor{lightgray}
									 &  &  &  &  3.6052$\times10^{-5}$ & 7729.0 & 4237.3\\
\rowcolor{lightgray}
									 &  &  &  &  5.4263$\times10^{-5}$ & 9748.6 & 5341.6\\
\rowcolor{lightgray}
									 &  &  &  &  7.2203$\times10^{-5}$ & 11430.6 & 6263.1\\
\rowcolor{lightgray}
									 &  &  &  &  9.0143$\times10^{-5}$ & 12900.4 & 7076.4\\
\rowcolor{lightgray}
									 &  &  &  &  9.0062$\times10^{-6}$ & 3331.6 & 1857.7\\
\rowcolor{lightgray}
							&&&	933	&	4.8592$\times10^{-6}$ & 4971.5 & 2587.5\\
\rowcolor{lightgray}
									 &  &  &  &  9.0963$\times10^{-6}$ & 7974.8 & 4132.0\\
\rowcolor{lightgray}
									 &  &  &  &  1.8382$\times10^{-5}$ & 12960.3 & 6681.2\\
\rowcolor{lightgray}
									 &  &  &  &  3.6052$\times10^{-5}$ & 20041.0 & 10328.2\\
\rowcolor{lightgray}
									 &  &  &  &  5.4263$\times10^{-5}$ & 25483.4 & 13150.8\\
\rowcolor{lightgray}
									 &  &  &  &  7.2293$\times10^{-5}$ & 29999.1 & 15500.0\\
\rowcolor{lightgray}
									 &  &  &  &  9.0323$\times10^{-5}$ & 33904.5 & 17533.3\\
\verb+TT03+ &	182	& 20.1&	361	&	4.4986$\times10^{-6}$ & 2181.9 & 1333.9\\
									 &  &  &  &  9.0062$\times10^{-6}$ & 3691.9 & 2141.1\\
									 &  &  &  &  1.7931$\times10^{-5}$ & 5837.8 & 3283.4\\
									 &  &  &  &  3.6052$\times10^{-5}$ & 8967.5 & 4953.9\\
									 &  &  &  &  5.3992$\times10^{-5}$ & 11283.0 & 6203.1\\
									 &  &  &  &  7.2203$\times10^{-5}$ & 13179.8 & 7238.5\\
									 &  &  &  &  9.0233$\times10^{-5}$ & 14748.9 & 8100.2\\
									 &  &  &  &  9.0963$\times10^{-6}$ & 3675.2 & 2113.2\\
							&&&	1121	&	4.8592$\times10^{-6}$ & 5236.3 & 2718.4\\
 &  &  &  &  9.0963$\times10^{-6}$ & 8911.4 & 4605.2\\
 &  &  &  &  1.8382$\times10^{-5}$ & 14431.7 & 7426.0\\
 &  &  &  &  3.6052$\times10^{-5}$ & 22734.8 & 11680.6\\
 &  &  &  &  5.4263$\times10^{-5}$ & 29132.4 & 14987.9\\
 &  &  &  &  7.2293$\times10^{-5}$ & 34556.9 & 17795.4\\
 &  &  &  &  9.0323$\times10^{-5}$ & 39163.0 & 20188.6	\\
\rowcolor{lightgray} 
\verb+TT04+ &	181	& 20.9 &	297	&	4.5887$\times10^{-6}$ & 2168.5 & 1260.1\\
\rowcolor{lightgray}
 &  &  &  &  9.0062$\times10^{-6}$ & 3477.7 & 1962.3\\
\rowcolor{lightgray}
 &  &  &  &  1.8021$\times10^{-5}$ & 5447.9 & 3015.6\\
\rowcolor{lightgray}
 &  &  &  &  3.6142$\times10^{-5}$ & 8245.4 & 4517.3\\
\rowcolor{lightgray}
 &  &  &  &  5.4082$\times10^{-5}$ & 10368.2 & 5673.6\\
\rowcolor{lightgray}
 &  &  &  &  7.2293$\times10^{-5}$ & 12171.4 & 6659.3\\
\rowcolor{lightgray}
 &  &  &  &  9.0143$\times10^{-5}$ & 13750.9 & 7520.2\\
\rowcolor{lightgray}
 &  &  &  &  9.0963$\times10^{-6}$ & 3513.9 & 1965.7	\\
\rowcolor{lightgray}
							&&&	723	&	4.5887$\times10^{-6}$ & 4362.3 & 2285.0\\
\rowcolor{lightgray}
 &  &  &  &  9.0963$\times10^{-6}$ & 7277.7 & 3786.0\\
\rowcolor{lightgray}
 &  &  &  &  1.8021$\times10^{-5}$ & 11648.5 & 6031.3\\
\rowcolor{lightgray}
 &  &  &  &  3.6052$\times10^{-5}$ & 17999.3 & 9317.5\\
\rowcolor{lightgray}
 &  &  &  &  5.4082$\times10^{-5}$ & 22850.9 & 11838.6\\
\rowcolor{lightgray}
 &  &  &  &  7.2113$\times10^{-5}$ & 26698.1 & 13862.7\\
\rowcolor{lightgray}
 &  &  &  &  9.0233$\times10^{-5}$ & 29833.5 & 15518.5	\\
\verb+TT05+ &	189	& 19.9 &	391	&	4.4986$\times10^{-6}$ & 2514.8 & 1487.8\\
 &  &  &  &  9.0963$\times10^{-6}$ & 4272.0 & 2411.7\\
 &  &  &  &  1.7931$\times10^{-5}$ & 6760.1 & 3725.0\\
 &  &  &  &  3.6142$\times10^{-5}$ & 10405.3 & 5648.9\\
 &  &  &  &  5.4172$\times10^{-5}$ & 13055.3 & 7069.4\\
 &  &  &  &  7.2203$\times10^{-5}$ & 15194.6 & 8223.3\\
 &  &  &  &  9.0143$\times10^{-5}$ & 16922.5 & 9168.9\\
 &  &  &  &  8.9160$\times10^{-6}$ & 4124.3 & 2327.3	\\
							&&&	823	&	4.4986$\times10^{-6}$ & 4659.8 & 2430.7\\
 &  &  &  &  9.0963$\times10^{-6}$ & 7878.7 & 4087.0\\
 &  &  &  &  1.8112$\times10^{-5}$ & 12693.8 & 6559.3\\
 &  &  &  &  3.6052$\times10^{-5}$ & 19651.8 & 10147.2\\
 &  &  &  &  5.3992$\times10^{-5}$ & 24961.4 & 12904.7\\
 &  &  &  &  7.2203$\times10^{-5}$ & 29391.6 & 15223.8\\
 &  &  &  &  9.0233$\times10^{-5}$ & 33089.3 & 17150.2	\\
\rowcolor{lightgray} 
\verb+TT06+ &	178	& 20.1 & 297	&	4.5887$\times10^{-6}$ & 1988.6 & 1188.7\\
\rowcolor{lightgray}
 &  &  &  &  9.0963$\times10^{-6}$ & 3218.3 & 1846.0\\
\rowcolor{lightgray}
 &  &  &  &  1.7931$\times10^{-5}$ & 5004.8 & 2806.9\\
\rowcolor{lightgray}
 &  &  &  &  3.6142$\times10^{-5}$ & 7604.3 & 4218.6\\
\rowcolor{lightgray}
 &  &  &  &  5.4082$\times10^{-5}$ & 9584.9 & 5297.6\\
\rowcolor{lightgray}
 &  &  &  &  7.2203$\times10^{-5}$ & 11254.4 & 6226.7\\
\rowcolor{lightgray}
 &  &  &  &  9.0233$\times10^{-5}$ & 12724.6 & 7040.0\\
\rowcolor{lightgray}
 &  &  &  &  9.0062$\times10^{-6}$ & 3215.6 & 1841.7	\\
\rowcolor{lightgray}				
							&&&	523	&	4.4986$\times10^{-6}$ & 3229.0 & 1718.6\\
\rowcolor{lightgray}
 &  &  &  &  9.0963$\times10^{-6}$ & 5345.4 & 2817.8\\
\rowcolor{lightgray}
 &  &  &  &  1.8112$\times10^{-5}$ & 8446.4 & 4441.4\\
\rowcolor{lightgray}
 &  &  &  &  3.6052$\times10^{-5}$ & 12907.8 & 6775.0\\
\rowcolor{lightgray}
 &  &  &  &  5.4082$\times10^{-5}$ & 16284.9 & 8557.9\\
\rowcolor{lightgray}
 &  &  &  &  7.2113$\times10^{-5}$ & 18945.8 & 9976.4\\
\rowcolor{lightgray}
 &  &  &  &  9.0143$\times10^{-5}$ & 21042.6 & 11101.6	\\

\verb+TT07+ &	186	& 20.1 & 310	&	4.4986$\times10^{-6}$ & 2285.2 & 1328.5\\
 &  &  &  &  9.0963$\times10^{-6}$ & 3722.8 & 2095.9\\
 &  &  &  &  1.8112$\times10^{-5}$ & 5837.6 & 3224.0\\
 &  &  &  &  3.6052$\times10^{-5}$ & 8796.6 & 4819.7\\
 &  &  &  &  5.4172$\times10^{-5}$ & 11081.8 & 6063.9\\
 &  &  &  &  7.2113$\times10^{-5}$ & 12973.0 & 7095.0\\
 &  &  &  &  9.0143$\times10^{-5}$ & 14628.7 & 8005.8\\
 &  &  &  &  9.0963$\times10^{-6}$ & 3691.9 & 2083.8	\\
							&&&	503	&	4.4986$\times10^{-6}$ & 3351.9 & 1785.7\\
 &  &  &  &  9.0963$\times10^{-6}$ & 5498.7 & 2899.2\\
 &  &  &  &  1.8021$\times10^{-5}$ & 8649.9 & 4548.8\\
 &  &  &  &  3.6052$\times10^{-5}$ & 13216.1 & 6935.2\\
 &  &  &  &  5.4082$\times10^{-5}$ & 16646.8 & 8745.7\\
 &  &  &  &  7.2203$\times10^{-5}$ & 19406.2 & 10212.9\\
 &  &  &  &  9.0143$\times10^{-5}$ & 21500.3 & 11350.6\\
 &  &  &  &  9.0062$\times10^{-6}$ & 5375.9 & 2834.0	\\
							&&&	940	&	4.4986$\times10^{-6}$ & 4396.8 & 2906.0\\
 &  &  &  &  9.0062$\times10^{-6}$ & 7827.0 & 4657.2\\
 &  &  &  &  1.8021$\times10^{-5}$ & 13051.0 & 7335.7\\
 &  &  &  &  3.6142$\times10^{-5}$ & 20724.7 & 11281.6\\
 &  &  &  &  5.4082$\times10^{-5}$ & 26584.4 & 14334.5\\
 &  &  &  &  7.2203$\times10^{-5}$ & 31520.7 & 16890.7\\
 &  &  &  &  9.0233$\times10^{-5}$ & 35631.2 & 19037.4\\
 &  &  &  &  9.0963$\times10^{-6}$ & 7828.1 & 4646.5	\\
							&&&	1320&	4.4986$\times10^{-6}$ & 5022.3 & 3672.2\\
 &  &  &  &  1.8021$\times10^{-5}$ & 16360.7 & 9465.3\\
 &  &  &  &  3.6142$\times10^{-5}$ & 26408.7 & 14619.9\\
 &  &  &  &  7.2203$\times10^{-5}$ & 40950.1 & 22123.8\\
 &  &  &  &  9.0143$\times10^{-5}$ & 46687.6 & 25101.3\\

\rowcolor{lightgray} 
\verb+TT08+ &	177	& 20.1 &	298	&	4.4084$\times10^{-6}$ & 1650.6 & 1134.7\\
\rowcolor{lightgray} 
 &  &  &  &  9.0062$\times10^{-6}$ & 2839.2 & 1795.3\\
\rowcolor{lightgray} 
 &  &  &  &  1.8112$\times10^{-5}$ & 4589.6 & 2745.6\\
\rowcolor{lightgray} 
 &  &  &  &  3.6142$\times10^{-5}$ & 7074.5 & 4097.4\\
\rowcolor{lightgray} 
 &  &  &  &  9.0062$\times10^{-6}$ & 2821.6 & 1799.4	\\
\rowcolor{lightgray}
							&&&	523	&	4.4084$\times10^{-6}$ & 2978.2 & 1612.6\\
\rowcolor{lightgray} 
 &  &  &  &  9.0963$\times10^{-6}$ & 4940.2 & 2633.3\\
\rowcolor{lightgray} 
 &  &  &  &  1.7931$\times10^{-5}$ & 7727.7 & 4094.3\\
\rowcolor{lightgray} 
 &  &  &  &  3.6052$\times10^{-5}$ & 11832.8 & 6251.7\\
\rowcolor{lightgray} 
 &  &  &  &  9.0062$\times10^{-6}$ & 4880.7 & 2608.9	\\
\rowcolor{lightgray}
							&&&	731	&	4.4084$\times10^{-6}$ & 2913.3 & 2394.6\\
\rowcolor{lightgray} 
 &  &  &  &  9.0062$\times10^{-6}$ & 5434.2 & 3706.6\\
\rowcolor{lightgray} 
 &  &  &  &  1.8112$\times10^{-5}$ & 9219.8 & 5672.0\\
\rowcolor{lightgray} 
 &  &  &  &  3.6052$\times10^{-5}$ & 14601.5 & 8464.0\\
\rowcolor{lightgray} 
 &  &  &  &  9.0062$\times10^{-6}$ & 5393.3 & 3688.3	\\
\rowcolor{lightgray}
							&&&	935	&	 4.4986$\times10^{-6}$ & 3529.4 & 2758.9\\
\rowcolor{lightgray} 
 &  &  &  &  9.0062$\times10^{-6}$ & 6518.2 & 4303.6\\
\rowcolor{lightgray} 
 &  &  &  &  1.8112$\times10^{-5}$ & 11134.4 & 6678.8\\
\rowcolor{lightgray} 
 &  &  &  &  3.6052$\times10^{-5}$ & 17731.8 & 10072.0\\
\rowcolor{lightgray} 
 &  &  &  &  9.0062$\times10^{-6}$ & 6502.9 & 4284.4	\\
\rowcolor{lightgray}
							&&&	1130	&	 4.4986$\times10^{-6}$ & 4090.8 & 3124.7\\
\rowcolor{lightgray} 
 &  &  &  &  9.0963$\times10^{-6}$ & 7550.2 & 4899.8\\
\rowcolor{lightgray} 
 &  &  &  &  1.8112$\times10^{-5}$ & 12866.3 & 7622.9\\
\rowcolor{lightgray} 
 &  &  &  &  3.6052$\times10^{-5}$ & 20569.2 & 11585.7\\
\rowcolor{lightgray} 
 &  &  &  &  9.0062$\times10^{-6}$ & 7535.9 & 4885.1\\
\rowcolor{lightgray}
							&&&	1321	&	4.7690$\times10^{-6}$ & 4670.6 & 3563.0\\
\rowcolor{lightgray} 
 &  &  &  &  9.1865$\times10^{-6}$ & 8435.4 & 5490.2\\
\rowcolor{lightgray} 
 &  &  &  &  1.8112$\times10^{-5}$ & 14353.5 & 8515.6\\
\rowcolor{lightgray} 
 &  &  &  &  3.6052$\times10^{-5}$ & 23169.2 & 13047.5\\
\rowcolor{lightgray} 
 &  &  &  &  9.0062$\times10^{-6}$ & 8281.0 & 5431.7\\
\rowcolor{lightgray}
							&&&	1507	& 4.4986$\times10^{-6}$ & 4640.0 & 3733.4\\
\rowcolor{lightgray} 
 &  &  &  &  9.0963$\times10^{-6}$ & 9002.3 & 5963.3\\
\rowcolor{lightgray} 
 &  &  &  &  1.8112$\times10^{-5}$ & 15621.3 & 9337.9\\
\rowcolor{lightgray} 
 &  &  &  &  3.6052$\times10^{-5}$ & 25334.9  & 14333.5\\

\verb+TT09+ &	181	& 20.1& 297	&	4.4986$\times10^{-6}$ & 1769.3 & 1301.3\\
 &  &  &  &  9.0062$\times10^{-6}$ & 2988.4 & 2039.2\\
 &  &  &  &  1.8021$\times10^{-5}$ & 4777.8 & 3160.3\\
 &  &  &  &  3.6052$\times10^{-5}$ & 7199.9 & 4744.5\\
 &  &  &  &  5.3992$\times10^{-5}$ & 9047.6 & 6000.7\\
 &  &  &  &  7.2113$\times10^{-5}$ & 10610.6 & 7082.8\\
 &  &  &  &  9.0062$\times10^{-6}$ & 2977.5 & 2033.3\\
							&&&	1421	&	4.4986$\times10^{-6}$ & 6148.0 & 3296.5\\
 &  &  &  &  9.0963$\times10^{-6}$ & 10527.9 & 5625.8\\
 &  &  &  &  1.8112$\times10^{-5}$ & 17153.0 & 9173.7\\
 &  &  &  &  3.6322$\times10^{-5}$ & 26936.1 & 14498.7\\
 &  &  &  &  5.3992$\times10^{-5}$ & 34374.4 & 18590.4\\
\rowcolor{lightgray}
\verb+IP01+\tnote{1}   &     186    & 20.0 &   298     &      4.4986$\times 10^{-6}$ & 1311.7 & 849.3\\
\rowcolor{lightgray}
 &  &  &  &  9.0062$\times 10^{-6}$ & 2198.9 & 1353.0\\
\rowcolor{lightgray}
 &  &  &  &  1.8021$\times 10^{-5}$ & 3479.5 & 2092.2\\
\rowcolor{lightgray}
 &  &  &  &  3.6052$\times 10^{-5}$ & 5323.7 & 3162.3\\
\rowcolor{lightgray}
 &  &  &  &  5.4172$\times 10^{-5}$ & 6736.8 & 3999.9\\
\rowcolor{lightgray}
 &  &  &  &  7.2203$\times 10^{-5}$ & 7919.6 & 4703.8\\
\rowcolor{lightgray}
 &  &  &  &  9.0143$\times 10^{-5}$ & 8943.2 & 5332.0\\
\rowcolor{lightgray}
							&&&   723   &     4.4986$\times 10^{-6}$ & 2344.9 & 1422.3\\
\rowcolor{lightgray}
 &  &  &  & 9.0963$\times 10^{-6}$ & 4202.9 & 2407.1\\
\rowcolor{lightgray}
 &  &  &  & 1.8021$\times 10^{-5}$ & 7013.7 & 3906.8\\
\rowcolor{lightgray}
 &  &  &  & 3.5962$\times 10^{-5}$ & 11183.0 & 6131.0\\
\rowcolor{lightgray}
 &  &  &  & 5.4082$\times 10^{-5}$ & 14350.8 & 7839.5\\
\rowcolor{lightgray}
 &  &  &  & 7.2113$\times 10^{-5}$ & 16844.9 & 9202.2\\
\rowcolor{lightgray}
 &  &  &  & 9.0143$\times 10^{-5}$ & 18826.2 & 10299.7\\
\rowcolor{lightgray}
							&&&   1123 &     4.4986$\times 10^{-6}$ & 	2903.2 & 1856.8\\
\rowcolor{lightgray}
 &  &  &  & 9.0963$\times 10^{-6}$ & 5353.8 & 3130.5\\
\rowcolor{lightgray}
 &  &  &  & 1.8112$\times 10^{-5}$ & 9192.6 & 5146.0\\
\rowcolor{lightgray}
 &  &  &  & 3.6052$\times 10^{-5}$ & 14944.7 & 8163.7\\
\rowcolor{lightgray}
 &  &  &  & 5.4172$\times 10^{-5}$ & 19478.5 & 10555.7\\
\rowcolor{lightgray}
 &  &  &  & 7.2203$\times 10^{-5}$ & 23235.5 & 12557.7\\
\rowcolor{lightgray}
 &  &  &  & 9.0233$\times 10^{-5}$ & 26422.2 & 14275.6\\
\rowcolor{lightgray}			
							&&&   1503 &    4.4896$\times 10^{-6}$ & 2975.4 & 2241.0\\
\rowcolor{lightgray}			
 &  &  &  &  8.9070$\times 10^{-6}$ & 5856.7 & 3746.6\\
\rowcolor{lightgray}			
 &  &  &  &  1.8103$\times 10^{-5}$ & 10674.1 & 6253.1\\
\rowcolor{lightgray}			
 &  &  &  &  3.6043$\times 10^{-5}$ & 17831.2 & 9975.6\\
\rowcolor{lightgray}			
 &  &  &  &  5.4073$\times 10^{-5}$ & 23516.1 & 12964.0\\
\rowcolor{lightgray}			
 &  &  &  &  7.2104$\times 10^{-5}$ & 28343.6 & 15514.3\\
\rowcolor{lightgray}			
 &  &  &  &  9.0134$\times 10^{-5}$ & 32542.2 & 17737.7\\

\end{supertabular}
\end{small}

\end{document}